\documentstyle[12pt,aasms4]{article}

\lefthead{Allende Prieto, Ruiz Cobo, Garc\'{\i}a L\'opez, Lambert \& Gustafsson}
\righthead{Semi-Empirical Model Atmospheres}
\begin{document}

\title{Model Photospheres for Late-Type Stars from the Inversion of
High-Resolution Spectroscopic Observations. Groombridge 1830 and
$\epsilon$ Eridani}

\author{Carlos Allende Prieto}
\affil{McDonald Observatory and Department of Astronomy, \\ The University of Texas at Austin \\ RLM 15.308, Austin, TX 78712-1083, \\ USA}

\author{Ram\'on J. Garc\'{\i}a L\'opez}
\affil{Instituto de Astrof\'\i sica de Canarias \\ E-38200, La Laguna,
Tenerife,  \\  SPAIN}

\author{David L. Lambert}
\affil{McDonald Observatory and Department of Astronomy, \\ The University of Texas at Austin \\ RLM 15.308, Austin, TX 78712-1083, \\ USA}

\author{Basilio Ruiz Cobo}
\affil{Instituto de Astrof\'\i sica de Canarias \\ E-38200, La Laguna,
Tenerife,  \\  SPAIN}

\authoraddr{IAC, E-38200, La Laguna, Tenerife,  SPAIN}

\authoremail{callende@astro.as.utexas.edu, rgl@iac.es, 
dll@astro.as.utexas.edu, brc@iac.es, }

\begin{abstract}

An  inversion technique to recover LTE one-dimensional model
photospheres for late-type stars,  which was previously applied to the
Sun (Allende Prieto et al. 1998), is now employed to reconstruct,
semi-empirically, the photospheres of cooler dwarfs: the metal-poor
Groombridge 1830 and  the active star of solar-metallicity $\epsilon$
Eridani. The model atmospheres we find reproduce satisfactorily
all the considered weak-to-moderate neutral lines of metals,  satisfying 
in detail the excitation equilibrium of iron, the wings of strong lines, 
and the slope of the optical continuum.
The retrieved models show a slightly steeper temperature
gradient than flux-constant model atmospheres in the layers   where
$\log \tau \le -0.5$.  We argue that these differences should reflect
missing ingredients in the flux-constant models and point to
granular-like inhomogeneities as the best candidate.

The iron ionization equilibrium is well satisfied by the model for 
Gmb1830, but not for $\epsilon$ Eri, for which a discrepancy of 0.2 
dex between the logarithmic iron abundance derived from neutral
and singly ionized lines may signal departures from LTE.
The chemical abundances of calcium, titanium,
chromium, and  iron derived with the empirical models from neutral lines 
do not differ much from previous analyses based on flux-constant
atmospheric structures.

\end{abstract}
 

\section{Introduction} 
\label{sec1}

The measurement of chemical abundances from stellar spectra 
 relies on a series of assumptions about the physical
properties of the stellar atmosphere. Ideally a model atmosphere should be
recoverable from the observed electromagnetic spectrum, but due to
observational limitations  models are
 either  constructed from a few physical principles, that lead to a closed
system of differential equations and boundary conditions,  or modeled
from some observed  spectral features  constrained by some 
theoretical bases. Such models are here referred to as theoretical and
empirical (or semi-empirical) model atmospheres, respectively.
The comparison between model atmospheres derived by different methods
can be used to test our actual knowledge on the   structure of
the stellar atmosphere. Good agreement exists between theoretical 
and empirical models for the
temperature stratification of the solar photosphere. Unlike the solar case,
where the high quality of the spectroscopic observations has motivated
both the empirical modeling and  theoretical studies of its atmosphere,
the analyses of more distant late-type
 stars are commonly carried out using
relatively simple  theoretical models for their photospheres.
As an example, it is rare to find studies in the literature analyzing in detail the
likely errors that occur when interpreting the stellar spectra with model
atmospheres based on approximations such as Local Thermodynamical
Equilibrium (LTE).

Previous efforts to  model empirically the photospheres and
chromospheres of cool stars others than the Sun, such as those by
M\"ackle et al. (1975) for Arcturus, Ruland et al. (1980) for Pollux,
Magain (1985) for the metal poor sub-giant HD140283, or Thatcher et
al.  (1991)  for $\epsilon$ Eri were severely limited by the quality of
the spectroscopic observations. 
Technical advances in  astronomical instrumentation have
 made it possible to acquire data more comparable to that for
 the Sun. Extremely high resolving power and
signal-to-noise ratio is feasible for many stars, at
least  to  seventh magnitude.  In this environment, 
we have  reconsidered the possibility of  semi-empirical modeling
 the photospheres of cool stars by developing
an inversion code of stellar spectra. The method has been previously
tested with the Sun (see Allende Prieto et al. 1998), demonstrating
that it is able to recover the depth-stratification of the solar
photosphere from {\bf normalized} spectral line profiles. The procedure
involves the assumption that the stellar photosphere is plane-parallel,
in LTE, in  steady state, and in hydrostatic  equilibrium. The star is
assumed to rotate as a solid body. Magnetic fields are
neglected.

We have selected two well-known nearby stars for this study: the
metal-poor G8 dwarf Gmb1830 (HD103095; HR4550; [Fe/H]\footnote{[M/H] =
$\log (\frac{N({\rm M})}{N({\rm H})}) - \log (\frac{N({\rm M})}{N({\rm
H})})_{\odot}$  where $N$(M) is the number density of the nuclei of the
element M and  "H" refers to hydrogen.} $\sim$ --1.3) and the
solar-like metallicity K2V $\epsilon$ Eri (HD22049; HR1084).  Gmb1830
is  the brightest star ($V=6.42$) that is significantly  metal
deficient. It has been studied widely  making use of theoretical model
 atmospheres and high resolution spectroscopic observations, e.g.
 Smith, Lambert, \& Ruck (1992) and Balachandran \& Carney (1996).  The
star was reported to show radial velocity variations (Beardsley,
Gatewood, \& Kamper 1974), but  subsequent detailed studies (Griffin
1984; Heintz 1984) did not confirm the variations.
 The star shows a periodic variation of the emission in the Ca II H and
K lines, likely reflecting  a solar-like activity cycle with a period
of about 7 years (Wilson 1978; Radick et al. 1998). $\epsilon$ Eri is a
young and active dwarf  surrounded by a
 ring of dust at a distance of   60 AU (Greaves et al. 1998). Its line
bisectors, magnetic activity, and temperature have  been observed to
vary by Gray \& Baliunas (1995). 


We have obtained high-quality spectroscopic data for these stars and
followed  the inversion procedure 
 previously applied to the Sun.  Next section
describes the observations and the database employed in the study. \S3
describes the details of the inversion procedure, and \S4 
the retrieved model atmospheres and their comparison  with  
observations, while \S5  discusses and summarizes the main 
conclusions.

\section{Observations of high resolution line profiles. Archived and
previously published complementary spectroscopic information}
\label{sec2}

Optical spectroscopic observations
were carried out in  1996 February using the higher resolution camera of
the  {\it 2dcoud\'e} echelle spectrograph (Tull et al. 1995) coupled to the
Harlan J. Smith Telescope at  McDonald Observatory (Mt.
Locke, Texas). The cross-disperser and the availability of a $2048 \times  
2048$ pixels CCD detector made it possible to gather up to 300 \AA\ in a
single exposure. The set-up
provided a resolving power of  
$ \lambda/\Delta\lambda \simeq$ 200000.  As many 1/2 hour
exposures were acquired as were needed to reach a signal to noise ratio
(SNR) of $\sim$ 300--800. Table 1 describes the observational program.

A very careful data reduction was applied using the
 IRAF\footnote{IRAF is distributed
 by the National Optical Astronomy Observatories, which are 
operated by the Association of Universities for Research in 
Astronomy, Inc., under cooperative agreement with the 
National Science Foundation.} software
package, and consisted of: overscan (bias) and scattered light
subtraction, flatfielding, extraction of one-dimensional spectra,
wavelength calibration and continuum normalization. Wavelength
calibration was performed for each individual image using
$\sim$ 300 Th-Ar lines spread over the detector. The possibility of
acquiring daylight spectra with the same spectrograph allowed us to
perform a few interesting tests. Comparison of the wavelengths of 60
lines in a single daylight spectra (SNR $\sim$ 400-600, depending on
the spectral order)  with the high accuracy wavelengths measured in the
solar flux spectrum by Allende Prieto \& Garc\'{\i}a L\'opez (1998)
showed that the rms difference was at the level of 58 m s$^{-1}$
($\sim \frac{1}{11}$ pixel).  Before co-adding the individual
one-dimensional spectra, they were first cross-correlated to correct
for the change in Doppler shifts and instrumental drifts. More details are
given in Allende Prieto et al. (1999a).

Measurements of the 
 optical continuum flux are available  for the two stars, although the
fluxes are not on an absolute scale. The Breger (1976) catalogue
includes both stars, and Gmb1830 was also observed by
Peterson \& Carney (1979) and Carney (1983). The IUE\footnote{The
spectra were retrieved from the IUE Final Archive; Garhart et al. 1997.}
satellite observed both stars, and their UV fluxes are on an absolute
scale, providing complementary information, but as $\epsilon$ Eri is a
chromospherically active star, its UV spectrum rich in emission
lines  is not adequate to study the star's photosphere. The 
 available spectra of Gmb1830 (Table 2)  covering wavelengths redder 
than 2000 \AA\ were critically
compared, and averaged.  The velocity shifts between individual
spectra were found to be smaller than $\sim$ 1 \AA,  unimportant so
for the analysis of the continuum.    Interstellar extinction was considered
 negligible.

\section{Inversion}
\label{sec3}

\subsection{Trigonometric gravities and initial metallicities}

The inversion code assumes hydrostatic equilibrium.
Therefore, gravity must be known to attempt the inversion. The chemical
abundances of the elements responsible for the atomic lines we use as
input data are derived in the inversion process, but an initial
appraisal of the overall chemical composition is  necessary.

Following a procedure strictly identical to that applied by Allende
Prieto et al.  (1999b) to more than two hundred cool stars of different
metallicities, we have derived the {\it trigonometric} gravities for
the two nearby stars studied here from the {\it Hipparcos}
parallaxes,  finding $\log g = 4.68 \pm 0.07$ dex for Gmb1830 and $4.84
\pm 0.07$ dex for $\epsilon$ Eri.  Spectroscopic
studies assign to Gmb1830  metallicities in the range --1.2 $\le$
[Fe/H] $\le$ --1.4 (see Smith et al. 1992). Analyses 
of $\epsilon$ Eri point to slightly lower than solar
metallicities, typically within --0.2 $\le$  [Fe/H] $\le$ 0.0 (Drake \&
Smith 1993).

\subsection{Input data and  inversion procedure}

The line profiles entering the inversion code MISS (Multi-line
Inversion of Stellar Spectra)
 were carefully selected following the same criteria employed for the
 solar case (Allende Prieto et al. 1998):  they should be included in
the compilation of solar lines by Meylan et al. (1993), their
transition probabilities should have been measured by Blackwell and
collaborators in Oxford (e.g.  Blackwell \& Shallis 1979), and they
should be weaker than 80 m\AA\ in equivalent width (W$_{\lambda}$), to
minimize both departures from LTE, and the underestimate of the line
damping by using the  Uns\"old approximation (see, e.g., Ryan 1998).
These criteria provided 13 lines of iron, calcium, titanium, and
chromium
 in the spectral range covered for Gmb1830, and 10 of them were also
useful for $\epsilon$ Eri (the restriction of the lines'
W$_{\lambda}$ to be smaller than 80 m\AA\ was slightly relaxed for
$\epsilon$ Eri). This is
a significantly smaller number of lines than the set employed for the
solar inversion but, as we shall demonstrate below, the atmospheric
structure can still be derived with confidence. The wings of the Ca I
 6162 \AA\  line were included, as discussed for the solar inversion
by Allende Prieto et al. (1998), using the theoretical estimates of
Spielfiedel et al. (1991) for the damping  due to collisions with
hydrogen atoms.  The line data is listed in Table 3: all lines were
used for the modeling of Gmb1830, and those employed 
for $\epsilon$ Eri are identified with an asterisk.

The inversion proceeds analogously to the solar case, starting from an
isothermal model photosphere 
($T$= 5000 K),  increasing progressively the
number of nodes until either no significant improvement in the fit of
the line profiles is achieved or the temperature structure shows
wiggles, which are evidence that the degree of the chosen polynomial is
too high. The solar abundances (Anders \& Grevesse 1989)
  were taken as starting point
 for $\epsilon$ Eri,  while the abundances of the elements heavier than
helium were scaled by a factor 0.0316 ([M/H] = --1.5) for Gmb1830. The abundances assumed at the
beginning do not determine the final result. Fig. 1 shows the evolution
of the iron abundance through the inversion procedure 
for Gmb1830, assuming different initial guesses.  
An initial iron 
abundance that was more than $\sim$ 0.3 dex 
from [Fe/H] = -- 1.5  provided a
significantly different final abundance and a  model
photosphere that did not fit adequately the observed line profiles.   The
rotational velocity and the Gaussian macroturbulence were allowed to vary,
while the microturbulence was assumed to be solar ($\sim$ 0.6 km
s$^{-1}$),  negligible, or larger than 1  km s$^{-1}$, finally keeping the
best value from the point of view of the $\chi^2$ criterion between
observed and synthetic spectra. A Gaussian  profile was 
used to represent for the instrumental profile.

Uncertainties have been estimated following the algorithm described by
S\'anchez Almeida (1997):
\begin{equation}
\sigma^2_k=\alpha^{-1}_{kk} \chi^2/N
\end{equation}
where $\alpha_{kl}$ represents the standard covariance matrix, and 
$N$ is the number of free parameters.
That is, errors are evaluated as the standard least-square estimate (see, for
instance, Press et al 1988), augmented by the square root of the ratio
between the number of data and the number of free parameters. With this
correction, uncertainties are reliable even in the case 
the minimum of $\chi^2$  has not been reached.

\subsection{A test of the depth coverage using the solar spectrum}

We can make use of the previously studied solar case to estimate the
boundaries of  the depth coverage reached with the current set of
spectral lines. The solar inversion model was derived from the solar
line spectrum in the FTS atlas of Kurucz et al. (1984), with a higher
signal-to-noise ratio and spectral resolution than our stellar
spectra.  The McDonald day-light spectra  were acquired with the same
spectrograph and similar signal to noise as the stellar spectra, and
provide the possibility to carry out a  test of  the results obtained
with the McDonald setup.  Thus, we have repeated the application of the
MISS code to the solar spectrum,  using the  10 lines selected in
common for Gmb1830 and $\epsilon$ Eri extracted from the McDonald
day-light spectra.  Fig. 2 shows  the retrieved model (solid line:
error bars  are shown) compared with the  solar model obtained  from
the inversion of 40 lines in the Kurucz et al.  (1984) atlas  (dashed
line), as described in Allende Prieto et al.  (1998).   The smaller
sample of lines narrows the photospheric region covered. 
Nonetheless, the agreement is reasonable for $-2.5
\ge \log \tau \ge -0.5$, with differences smaller than 200 K
suggesting that the ten selected spectral lines  map approximately
this part of the  solar photosphere.

\subsection{Derived model photosphere}

The derived model stellar photospheres are shown in Fig. 3, and
compared with theoretical model photospheres from the grid by Kurucz
(1992), and the exact solution to the gray atmosphere
  ($T^4 = \frac{3}{4} T_{\rm eff}^4 [\tau +
q(\tau)]$) of the  assigned effective temperatures (see below).  The quality of
 the final fit to the observed line spectra is shown in Fig. 4 for the
 two stars.  There are marked differences between the temperature
stratification of the semi-empirical models and their purely theoretical
counterparts. While the star's gravity is  determined with high
accuracy from the trigonometric parallaxes measured by Hipparcos
(Nissen, H$\o$g \& Schuster 1997; Allende Prieto et al. 1999b) and some
additional hypotheses, the effective temperature
 typically gives discrepant values when derived from different methods
such as photometry, Balmer lines, excitation equilibrium, the spectral
energy distribution,  or temperature sensitive line ratios. We treat
the effective temperature of the appropriate theoretical model
 as an unknown parameter. Metallicity
  influences only weakly the considered spectral features.

The  derived rotational velocities  are 2.5 and 2.1  km s$^{-1}$ for 
Gmb1830 and $\epsilon$ Eri, respectively. They compare well
with those empirically measured by Fekel (1997): 2.2 and 2.0 km
s$^{-1}$. Mayor and collaborators  derived 1.5 km s$^{-1}$
 for $\epsilon$ Eri from the CORAVEL measurements (Benz \& Mayor 1984), while Gray (1984) gives 2.2 km s$^{-1}$. Smith et al. (1992)
 derived 1.8 km s$^{-1}$ for the combination of the different
 broadening mechanisms:  instrumental, macroturbulence and rotation.
However, our rotational velocities should be taken with caution.  The
inversion code is not able to  cleanly unravel the Gaussian
macroturbulence from  the rotational broadening profile.  Moreover, the
use of the Van der Waals approximation for the collisional broadening
with neutral hydrogen is expected to  underestimate, systematically,
the collisional broadening, and should produce  larger-than-real
estimates for the rotation-macroturbulence broadening. The derived
Gaussian macroturbulence is 0.0 and 1.5 km s$^{-1}$ for Gmb1830 
and $\epsilon$ Eri, respectively, and in both cases the preferred 
microturbulence was  0.6 km s$^{-1}$. 

Obviously, the   abundances obtained directly from the inversion
are only those of the elements whose lines are represented in the sample
selected as input data. These are: calcium, titanium, chromium, and
iron. The results appear in Table 4.
The derived ratio of iron to calcium
 abundances for Gmb1830 agrees very well with that found by Smith et
al. (1992)  making use of MARCS  model atmospheres (Gustafsson et al. 1975),    and Balachandran \&  Carney (1996) making use of those of Kurucz (1992).
But the iron abundance with respect to the Sun derived by both groups, 
is  $\sim$ 0.1 dex higher than
ours. The comparison of the abundances of these elements in $\epsilon$
Eri  with the determination by Drake  \& Smith (1993) shows a
discrepancy for calcium of $+0.13$ dex  (A(Ca) = 6.26 --
 6.39)\footnote{A(M) = $\log \left(\frac{N({\rm M})}{N({\rm H})}\right) + 12$.} 
and a difference   of  0.28 dex for iron:  our result is
 [Fe/H] = $+0.19$ dex, and theirs was [Fe/H] = $-0.09$. These and others 
inconsistencies found for $\epsilon$ Eri (see Drake \& Smith (1993) and 
\S 5 of this paper) might  be partly related to the magnetic activity of
the star.

\section{Spectroscopic properties of the derived semi-empirical models}
\label{sec4}

We have made use of several
 spectroscopic indicators to test the depth stratification of
the stellar photosphere: the optical spectral energy distribution, weak 
metal lines spanning a wide range in excitation potential, and collisionally
enhanced wings of strong metallic lines. The optical continuum and the
excitation balance of weak metal lines are highly sensitive to temperature.
The wings of the very few strong metal lines for which detailed theoretical
calculations or laboratory measurements of their damping constants is 
available (Lambert 1993) are reliable estimators of the pressure in the
line forming region (see, e.g., Edvardsson 1988, Anstee, O'Mara \& Ross 1997).
Other tools are available, such as the wings of the Balmer lines (Fuhrmann, 
Axer \& Gehren 1993), but have not been included here because they are more   complicate to interpret. The reader is referred to Fuhrmann et al. (1993) for an extensive discussion on the analysis of hydrogen lines.

\subsection{Stellar continuum. Optical and UV}

While most of the spectrophotometric measurements in the literature do
not provide an estimate of their accuracy, the availability of different
{\bf independent} determinations allows us to  derive empirically their
precision for the case of  Gmb1830. Figure 5a compares the observed
optical fluxes with the models' prediction normalized at 7500 \AA
($1/\lambda(\mu{\rm m})\simeq$ 1.33),  the reddest wavelength where
all  the different observational sources  have data. 
Independently observed
fluxes are represented by filled circles (Breger 1976), open circles
 (Peterson \& Carney 1979) and asterisks (Carney 1983). The true
 continuum (no line blanketing), given the low metallicity of the star,
is expected to fall very close to the observed continuum, except in the
blue part of the spectrum, where it should be higher, consistent with
the presence of many absorption lines. The prediction of the  MISS
model (solid line) shows this behavior. It is shown in
 the Figure that, for the fixed gravity and metallicity ($\log g$ =
4.68; [Fe/H] = --1.3), a theoretical model  atmosphere with an
effective temperature $T_{\rm eff} \simeq$ 5050 K reproduces the
observations.  This was already pointed out by
Balachandran \& Carney (1996). The fluxes for the theoretical (Kurucz 1992)
models\footnote{These fluxes were obtained through interpolation
 in the Kurucz's grid, available at the CCP7 
web site: http://ccp7.dur.ac.uk.} take into account the  
presence of lines (unlike the MISS continuum) 
and are therefore  directly comparable with the observations.
The effective temperature derived from the optical continuum
 is consistent, as expected, with that recently derived by 
Alonso, Arribas, \& Mart\'{\i}nez Roger (1996) making use 
of the  Infrared Flux  Method (IRFM;  Blackwell et  al. 1990): 5029 K.

The absolutely calibrated UV spectra of Gmb1830 in the IUE final
archive (IUEFA) offer us the possibility to carry out  an independent
test.  Combining the apparent brightness of the star in the Johnson V
band, V = 6.42, and the {\it Hipparcos} parallax, $\pi$ = 0.109 mas, we
arrive at an absolute magnitude for this star $M_V=6.61$.  Using this
value and the star's metallicity to choose an isochrone from
$\alpha$-elements enhanced models of
 Bergbusch \& Vandenberg (1992),  quite independently of the assumed
star's age  due to the fact that the star has not evolved from the main
sequence, we find the stellar mass to be M = 0.64  $\pm$ 0.05
M$_{\odot} $, in agreement with the older estimate by Smith et al.
(1992), and the stellar radius R = 0.61 $\pm$ 0.05  R$_{\odot}$.  The
radius and the parallax directly provide the dilution factor of the
flux as the light travels from the star to Earth,  making it possible to
compare models' surface fluxes and the IUE observations. Fig. 5b
re-enforces the conclusion previously obtained from the slope of the
optical continuum, that the $T_{\rm eff}$ of the theoretical model
atmosphere is close to 5050 K.  Unfortunately, at the present stage
we cannot carry out a detailed spectral synthesis, including 
the many lines present in this spectral range, with the MISS model. 
 However, it is unclear whether the  lines used  here for the modeling 
are able to constrain the layers of the photosphere where the UV 
continuum is forming.

The optical continuum of $\epsilon$ Eri, as appears in Breger's catalog, 
has been represented in Fig.  6. Again, the MISS model (solid line)
predicts a slope compatible with the observations. The  effective
temperature for a theoretical model that fits the continuum slope is
somewhat hotter than $T_{\rm eff} \simeq$  4850 K (dashed line), but
cooler than 5200 K.  Alonso et al.  (1996)  derived 5076 K, and this is the
temperature that we assign to the theoretical models shown in Fig. 3b.
 We recall that the chromospheric activity of $\epsilon$ Eri dominates
the UV spectrum of the star, excluding the possibility of studying the
photosphere from this spectral region.  Using the isochrones of
Bergbusch \& Vandenberg (1992) we find that $\epsilon$ 
Eri's mass is M =  0.76 $\pm 0.05$ M$_{\odot}$, 
and  its radius R = 0.55 $\pm$
0.05  R$_{\odot}$.

\subsection{Weak lines. Excitation equilibrium of Fe I}

The highly accurate determinations of the transition probabilities for a
large sample of neutral iron lines  by O'Brian et al. (1991) provide
an independent test of the semi-empirical model. We have identified 
12 iron lines in O'Brian et al.'s list within  our spectral range,  covering a 
significant range in excitation
 potential and equivalent width to explore the excitation equilibrium
of neutral iron for the considered model atmospheres. The lines are
listed in Table 5, with their measured equivalent widths.

The MISS model for Gmb1830,   with the derived solar-like microturbulence,
does not exhibit  a significant dependence of the derived iron abundance
on the equivalent width. The upper panel of Fig. 7 shows the
differences between the abundances observed and predicted by MISS, as
derived from the differences between  observed and predicted equivalent
widths.  The slope of the linear (least-squares) 
model is  $0.002 \pm 0.007$. 
The lower panel shows that the
excitation equilibrium is satisfied:  the slope of the derived abundance
against excitation potential is $0.012 \pm 0.074$.  

The Fe I lines in the O'Brian et al's list identified in the spectrum
of $\epsilon$ Eri are the same ones 
observed in Gmb1830, except for $\lambda$
5321 \AA. Of course,  given the higher metallicity, the lines are
stronger in this case. Figure 8 shows that the microturbulence 
retrieved in the modeling process for
$\epsilon$ Eri induces no significant gradient in the abundance as
derived from lines of different strength. The excitation equilibrium is
satisfied for this set of lines as well: the slope of the  abundance differences
 as a function of the excitation potential is $0.018 \pm 0.045$.

\subsection{Wings of strong metallic lines}

The wings of the Ca I $\lambda$ 6162 line were used as input for the
semi-empirical modeling, and 
the spectral region close to the line is very useful as
weak  calcium lines are present, allowing a test of the retrieved
model and calcium abundance.  
In Fig. 9 (upper panel) the observed spectrum is shown
(dots), and compared with the synthesis using the MISS structure (solid
line). The MISS model  reproduces nicely not only the observed wings of the
strong line, as imposed in the modeling process, but also the
surrounding calcium and iron lines, with the derived calcium
abundance: A(Ca) = 5.27, or [Ca/H] = --1.09, which is 0.3 dex higher
than the derived iron abundance. The result, that fully agrees with the
analyses using MARCS 
model atmospheres by Smith et al.  (1992), reflects the
well-known over-abundance of $\alpha$-elements 
in metal-poor stars. Departures from LTE are expected in the  core
of the $\lambda$ 6162 \AA\ line.

The lower panel of Fig. 9 shows the same spectral region
  for $\epsilon$ Eri. The observed spectrum (dots) is nicely reproduced
by the MISS model with [Ca/H] = $+0.03$. The oscillator strengths were
extracted from the Vienna Atomic Line Database (VALD), and have been
tested against the solar spectrum (Allende Prieto et al. 1998).

\section{Summary and conclusions}
\label{sec6}

We have applied an inversion method to normalized line profiles in the
optical spectra of the metal-poor dwarf Gmb1830 and the
solar-metallicity dwarf $\epsilon$ Eridani. This demonstrates the viability
of the empirical modeling to stars other than the Sun, to which the
inversion had been previously applied (Allende Prieto et al. 1998).
The semi-empirical models reproduce very well
weak-to-moderate lines of neutral atoms,  and satisfy the excitation
equilibrium of iron. 
The models also fit  the wings of strong lines, and 
the slope of the optical continuum.

The derived model atmospheres are slightly different from
 the theoretical models  of a similar effective temperature, showing a
 steeper temperature gradient. These differences must correspond to missing
ingredients in the theoretical modeling.  In our view, 
a likely  candidate  is stellar granulation. The
 semi-empirical models are one-dimensional, static, and
 time-independent too, but  flux-constancy is no longer imposed. 
This flexibility provides  room for
 missing physics in the complex dynamical interplay between matter
and  radiation. Therefore studying and analyzing the differences
between theoretical and semi-empirical structures  may help us to
recognize which physical effects are lacking. 
The mean temperature structure
derived from numerical simulations of solar granulation (Stein \&
Nordlund 1998) has shown a steeper  gradient for the layers
outwards than $\log \tau \simeq -0.5$,  resembling the behavior
of the semi-empirical models presented here. This effect was
 also apparent in the semi-empirical model  for the solar photosphere  
we derived using the same technique.

The differences between the semi-empirical model and the flux-constant models for Gmb1830 do not affect significantly 
 the abundances previously published for this star.
It is of interest that the absolute abundance of Li measured by
Deliyannis et al. (1994) in Gmb1830, namely  A(Li) = 0.27 dex, does not
change by more than 0.01 dex when using the semi-empirical model for
this star.

If departures from LTE are significant, the semi-empirical models would
adapt themselves to reproduce the line profiles under LTE. This effect
has been named  {\bf NLTE masking} and has been invoked to explain the
differences between the Holweger \& M\"uller (1974) empirical solar
model and solar NLTE models by Rutten \& Kostik (1982).  Quantifying
the importance of departures from LTE  should be performed
 through detailed calculation of  model atmospheres. Hauschildt, Allard
\& Baron  (1999) have already stepped forward to this, computing models
for the Sun and Vega, but these studies need to be extended to a wide
range of physical parameters. At this point, we can  take a glimpse at
the consistency of the results provided by the inversion procedure
checking the iron abundance that comes out from the analysis of ionized
iron lines. In late-type dwarfs, such as those analyzed here, most of
the iron is in form of Fe$^{+}$ ions and therefore, departures from LTE
ionization equilibrium are unlikely to disturb  the abundances derived
from lines of this species. Smith et al. (1992) and Drake \& Smith
(1993) analyzed four and three Fe II lines in the spectra of Gmb1830
and $\epsilon$ Eri, respectively. Using their atomic data, we
synthesized the lines with the semi-empirical models and the abundances
retrieved from  neutral lines, finding that the agreement between
observed and predicted equivalent widths for Gmb1830 is excellent,
always better than 1 m\AA. Conversely, the equivalent width predicted
for the Fe II lines of $\epsilon$ Eri, are systematically smaller than
the observations, leading to a higher iron abundance by 0.2 dex than
the Fe I lines, which may be an indicator of departures from LTE (we
recall that this star exhibits magnetic activity).  
It is worthwhile to note that Feltzing \& Gustafsson (1998) 
found  further evidence of overionization  (compared to LTE predictions)  
for several  K dwarfs.
Socas-Navarro, Ruiz Cobo, \& Trujillo Bueno (1998) 
have developed a NLTE inversion
procedure oriented to the study of the solar chromosphere. The
implementation of the method to stars is highly desirable, and  its
application to $\epsilon$ Eri may bring into agreement the abundances of
neutral and ionized lines.

Understanding of the atmospheric
structure and the line formation in metal-poor stars is of particular
relevance.  Detailed
abundance analyses on these stars provide precious information on the
chemical evolution of the Galaxy, how metals are
synthesized in stellar interiors, or
even the yields of the primordial nucleosynthesis. 
Very recently, Asplund et al. (1999) have computed the first
 hydrodynamical simulations of surface convection for metal-poor stars,
 similar to those of Stein \& Nordlund (1998) for the Sun. The mean
temperature structures they derive for
 HD140283 ([Fe/H] $\simeq-2.5$) and HD84937 ([Fe/H] $\simeq -2.3$)
 show  again a steeper gradient in the layers of $\log \tau \le -0.5$
 than the flux-constant stratification of the corresponding
 one-dimensional models. This turns out to have  important consequences for the  derived  lithium abundance, indicating that lithium abundances
could have been overestimated by $0.2-0.35$ dex in metal-poor stars
using one-dimensional model atmospheres.  Data
of similar quality to those presented in this paper, and even wider
spectral coverage have been collected for HD140283 during the past few
years (Allende Prieto et al.  1999a), and should provide an alternative
semi-empirical model for this star in the near future.

\acknowledgements

We thank  Martin Asplund, Luis Ram\'on Bellot Rubio,  Manolo Collados,
Klaus Fuhrmann, Bengt Gustafsson, and Nataliya Shchukina for fruitful
discussions. Suchitra Balachandran and Bruce Carney have kindly
provided measurements of the optical continuum  of Gmb1830, and
Benjam\'{\i}n Montesinos helped
 with the IUE data. We are grateful to the staff at McDonald
Observatory for their professional support. This research has been
partially supported by the NSF (grant AST-9618414), the Spanish DGES
(projects PB92-0434-C02-01 and PB95-1132-C02-01),
 and the Robert A. Welch Foundation of Houston, Texas.  NOS/Kitt Peak
FTS data used here were produced by NSF/NOAO.  We have made use of
VALD, the IUE final archive, data from the {\it Hipparcos} astrometric
mission of the ESA, and the SIMBAD database, operated at CDS
(Strasbourg, France).

\clearpage


\clearpage

\figcaption{Variation of the iron abundance 
through the inversion process for Gmb1830, starting from different
initial estimates.}

\figcaption{The solar MISS model atmosphere (Allende Prieto et al.
1998; dashed line), derived using 40 spectral lines in the Kurucz et
al. (1984) atlas,  compared with the model retrieved from
 the McDonald Observatory day-light spectrum
 using only the 10 lines in this study for Gmb1830 and $\epsilon$
Eri (solid line with error bars).}

\figcaption{a) The semi-empirical model for Gmb1830 (solid line), 
compared with  a model computed with ATLAS9 for 
$T_{\rm eff}$ = 5050 K (dashed) and  
$T^4 = \frac{3}{4} T_{\rm eff}^4 [\tau + q(\tau)]$  
(Gray atmosphere) 
b)  Semi-empirical and flux-constant models for $\epsilon$ Eri.} 

\figcaption{a) Fit of the observed line profiles (dots) in the spectrum of
Gmb1830 with the semi-empirical model (dashed line); b) the
same for the spectrum of $\epsilon$ Eri. The dots correspond to all the 
wavelengths which are actually fitted by the inversion procedure, 
joined together, regardless of the existence of 
gaps between the different segments corresponding to a single 
spectral line, or one of the wings of a spectral line.}

\figcaption{a) Normalized optical continuum of Gmb1830; data from Breger (1976): filled circles; Peterson \& Carney (1979): open circles; and
Carney (1983): asterisks. The {\it true} continuum predicted by the
MISS model (solid line) and the {\it blanketed} continua of the
Kurucz's models for $T_{\rm eff}$ = 4900, 5050, and 5300 K are also
shown. b) Near UV continuum of Gmb1830 observed by IUE (dots). 
The  continua of the Kurucz's models for $T_{\rm eff}$
 = 4900, 5050, and 5300 K are also shown.}

\figcaption{Normalized optical continuum of $\epsilon$ Eri; data 
from Breger (1976): filled circles. The {\it true} continuum predicted 
by the MISS model (solid line)
and the {\it blanketed} continua of the Kurucz's models for $T_{\rm eff}$ = 4650,  4850, and 5200 K are also shown.}

\figcaption{Differences between the abundances observed and predicted by
the MISS model for Gmb1830, as derived from the difference in 
the equivalent widths of Fe ~I lines in the O'Brian et al's list, against
observed equivalent widths (upper panel) and excitation potential (lower
panel). Also shown are linear regressions (least squares) indicating no
dependence of the abundances on these parameters.}

\figcaption{Same as Fig. 7, but for $\epsilon$ Eri.}

\figcaption{The region of the Ca ~I $\lambda$ 6162 \AA\ for  Gmb1830 (upper panel) and  $\epsilon$ Eri (lower panel),  
compared with the MISS model synthesis (solid line). 
The core of the Ca ~I $\lambda$ 6162 line is likely to be seriously affected
 by departures from LTE.}

\clearpage

\begin{deluxetable}{cccc}
\tablecaption{Observations: Dates, Spectral Ranges and Signal-to-noise Ratios. \label{table1}}
\tablehead{\colhead{Star} & \colhead{Spectral  range (\AA)} & \colhead{Date} 
& \colhead{SNR} }
\startdata 
$\epsilon$ \, Eri &	4851-6402 &	29-Feb-96 &	700 \\
"	 &	5204-7183 &	27/29-Feb-96 &	800 \\
Gmb1830 &	4853-6404 &	26-Feb-96 &	250-350 \\
"	 &	5204-7183 &	27/29-Feb-96 &	300-400 \\
\enddata
\end{deluxetable}


\begin{deluxetable}{ccc}
\footnotesize
\tablecaption{ Large Aperture Low Dispersion spectra available in the IUE Final
 Archive for Gmb1830 \label{table2}}
\tablehead{
\colhead{IUEFA Image} & \colhead{Acquisition Date} &  \colhead{Exposure time (s)}}
\startdata 
   LWR07351  &  30-Mar-80   &    120.000 \nl
   LWR07471  &  10-Apr-80   &    999.708 \nl
  LWP05802  &  23-Apr-85   &    359.506 \nl
   LWP13413  &  12-Jun-88   &    252.000 \nl
\enddata
\end{deluxetable}


\begin{deluxetable}{cccc}
\footnotesize
\tablecaption{ Spectral lines used in the inversion \label{table3}}
\tablehead{
\colhead{Element} & \colhead{Wavelength \tablenotemark{a}} & \colhead{Exc. Pot.} &  \colhead{$\log gf$}  \\
 & \colhead{(\AA)} & \colhead{(eV)} &  }
\startdata 
Ca ~I  & 6166.445$^{*}$   &  2.52  &--1.142  \nl
Ca ~I  & 6499.642$^{*}$   &  2.52  &--0.818  \nl
Ca ~I  & 6162.166$^{*}$   &  1.89  &--0.097  \nl
Ti ~I  & 5490.165   &  1.46  & --0.877  \nl
Ti ~I  & 6258.101$^{*}$   &  1.44  & --0.299  \nl
Cr ~I  & 5312.872$^{*}$   &  3.45  &--0.562  \nl
Cr ~I  & 5300.743$^{*}$   &  0.98  &--2.129  \nl
Fe ~I  & 5225.524$^{*}$   &  0.11  &--4.790  \nl
Fe ~I  & 5956.711$^{*}$   &  0.86  &--4.610  \nl
Fe ~I  & 6151.614$^{*}$   &  2.18  &--3.300  \nl
Fe ~I  & 6173.352   &  2.22  &--2.880  \nl
Fe ~I  & 6750.149$^{*}$   &  2.42  &--2.620  \nl
\enddata
\tablenotetext{a}{The lines used for modeling $\epsilon$ Eri are identified
with an asterisk}
\end{deluxetable}


\begin{deluxetable}{ccccc}
\footnotesize
\tablecaption{Abundances derived from the inversion \label{table4}}
\tablehead{
\colhead{Star} & \colhead{Element} & \colhead{Num. of lines} & \colhead{Abundance} & 
 \colhead{[M/H]}}
\startdata 
 {\it Gmb} 1830   &   Fe  &  7 &  6.09   &   $-1.39$  \nl
 "   &  Ca  &  3 & 5.27 &   $-1.09$  \nl
"   &   Ti & 3 & 3.78 & $-1.09$ \nl
 "   &  Cr & 2 & 4.26 & $-1.41$  \nl
 $\epsilon$ Eri   &   Fe  & 6 &   7.67   &   $+0.19$  \nl
 "   &  Ca  &  3 & 6.39 &   $+0.03$  \nl
"   &   Ti &  2 & 4.80  & $-0.19$ \nl
"   &  Cr & 2 & 5.73  & $+0.06$  \nl
\enddata
\end{deluxetable}


\begin{deluxetable}{rrrrr}
\footnotesize
\tablecaption{ Measured Fe I  lines  in the list of O'Brian et al. (1991) \label{table5}}
\tablehead{
 \colhead{Wavelength} & \colhead{Exc. Pot.} &  \colhead{$\log gf$}  & \colhead{W$_{\lambda}$ Gmb1830} & \colhead{W$_{\lambda} \epsilon$ Eri}\\
 \colhead{(\AA)} & \colhead{(eV)} &   &   \colhead{(m\AA)} &   \colhead{(m\AA)}}
\startdata 
5223.18   &  3.63  & $-1.78$ & 7.6  &  40.5  \nl
 5225.52   &  0.11  & $-4.76$ &  60.7  & 110.5  \nl
 5321.11   &  4.43  & $-1.09$  & 10.6   & \dots \nl
 5856.08   &  4.29  & $-1.33$  &  6.2  &  43.1 \nl
 5956.71   &  0.86  & $-4.50$  & 32.1  & 83.8 \nl
 6165.35   &  4.14  & $-1.47$  & 10.6 &   58.2  \nl
 5288.53   &  3.69  & $-1.51$  &  18.9  &  67.4 \nl
 5379.58   &  3.69  & $-1.51$  &  21.2  & 73.6 \nl
5464.28  &  4.14  &  $ -1.40  $ & 10.4  &  46.8 \nl
 6151.61   &  2.18  & $-3.37 $ &  23.4  &  72.8  \nl
 6498.95   &  0.96  & $-4.69$  &  25.8 & 78.1  \nl
6750.14   &  2.42  & $ -2.58$  & 44.6 &  103.1 \nl
\enddata
\end{deluxetable}

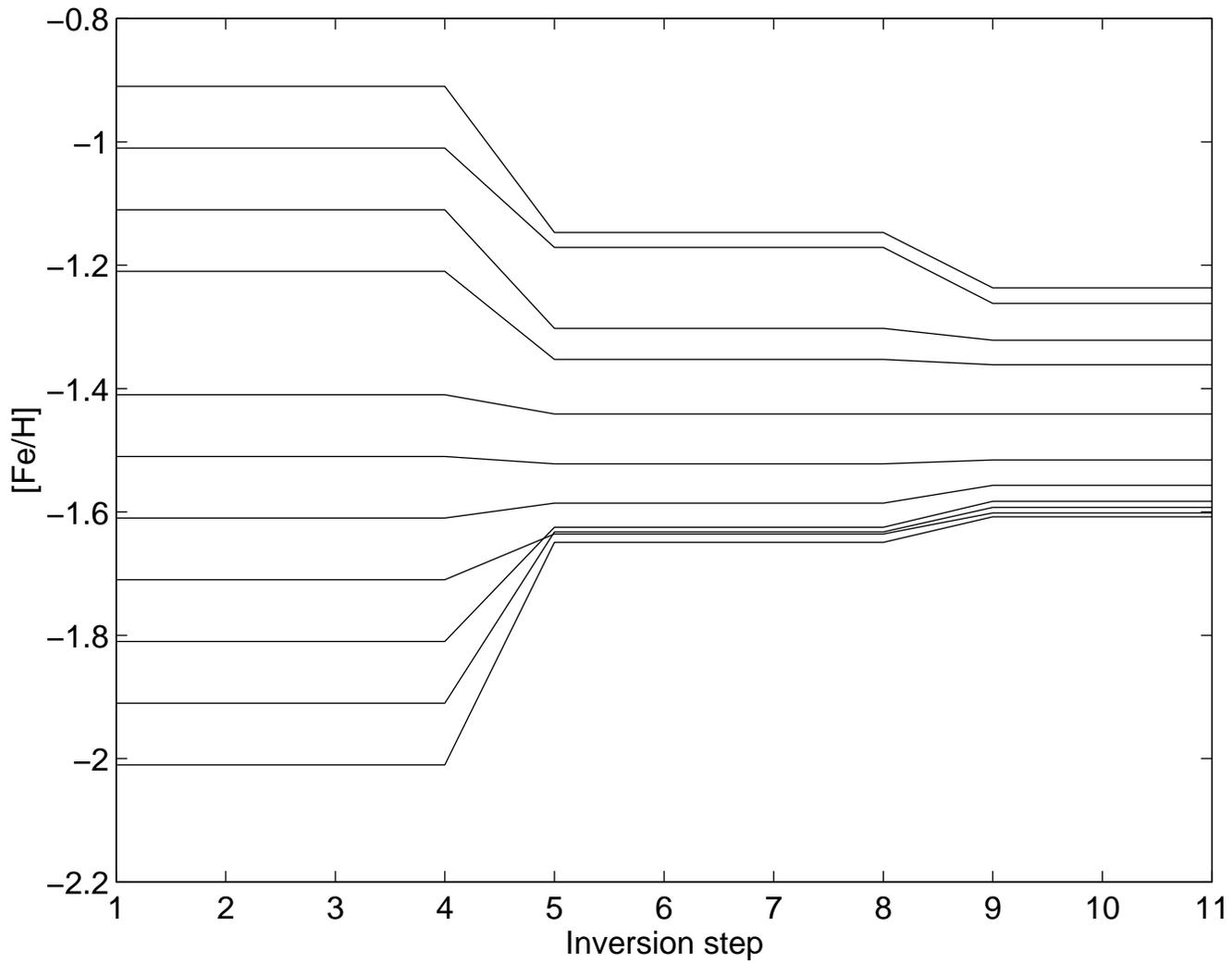
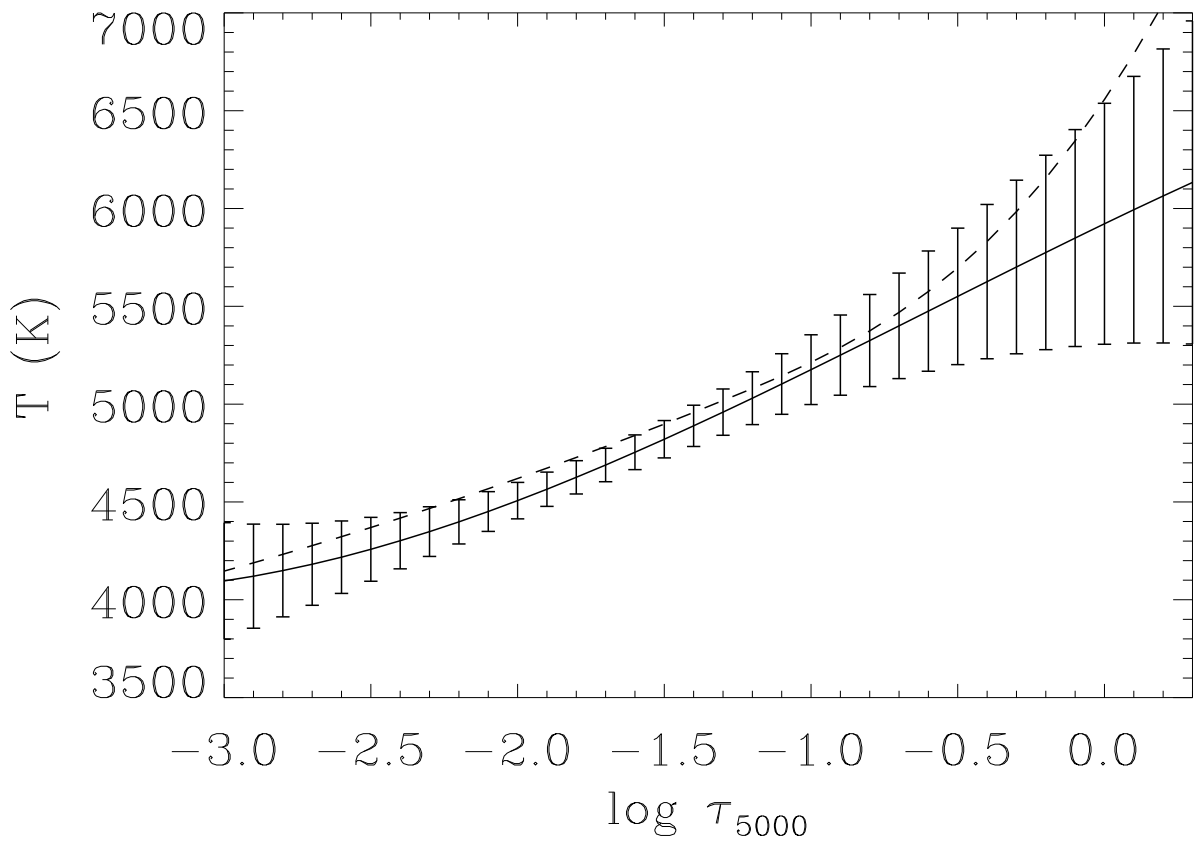
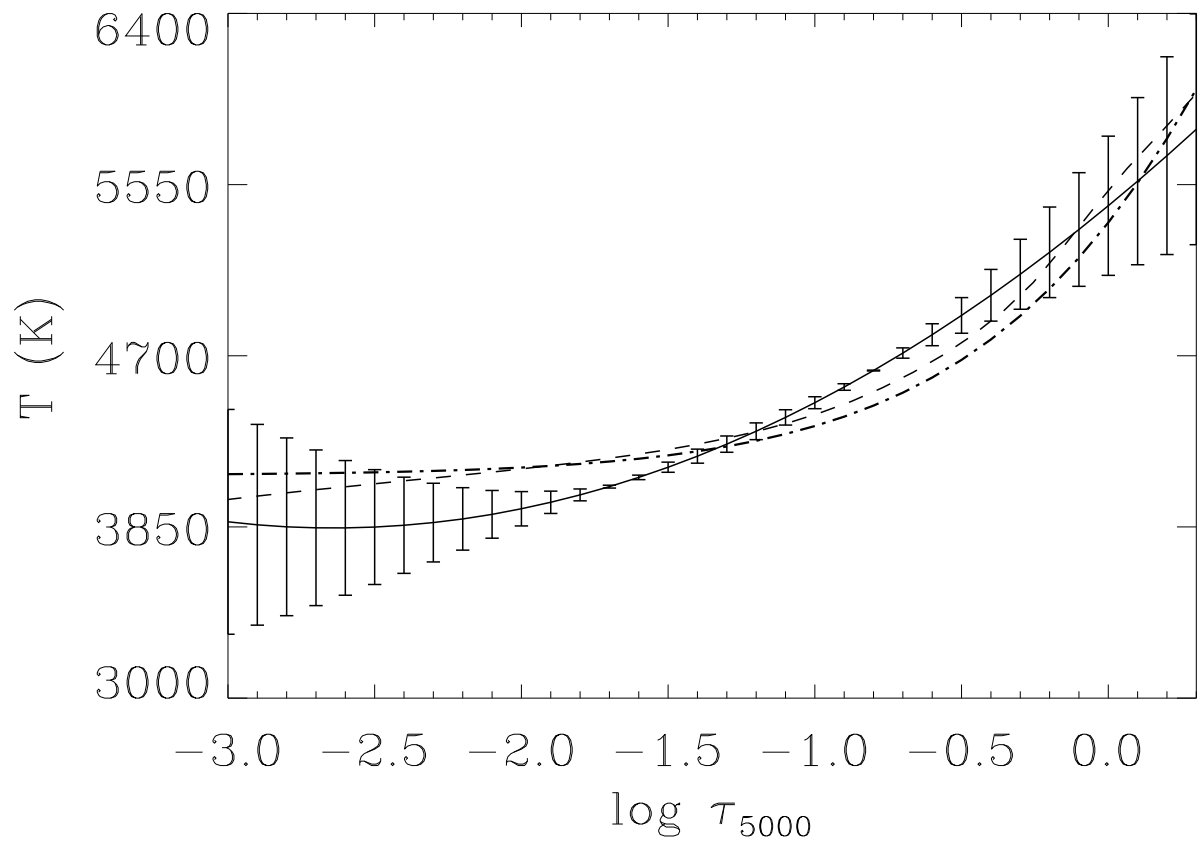
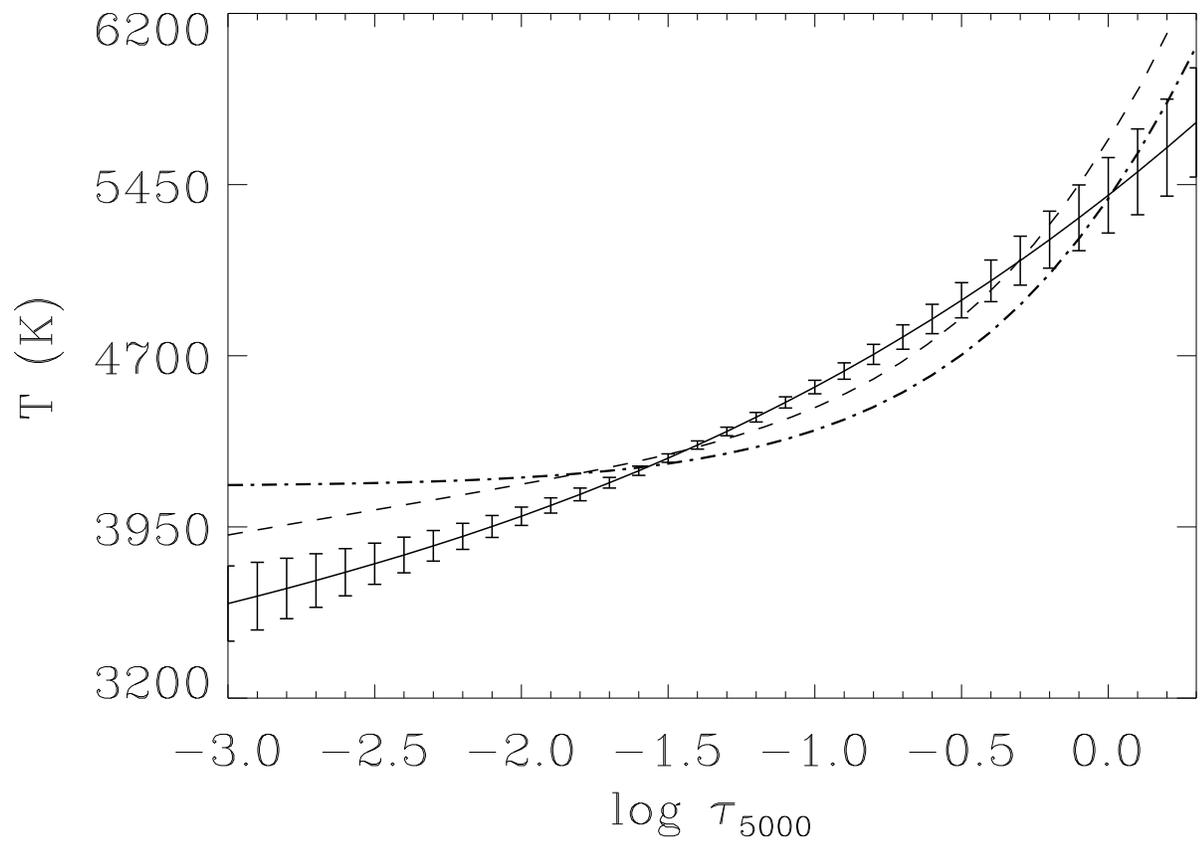
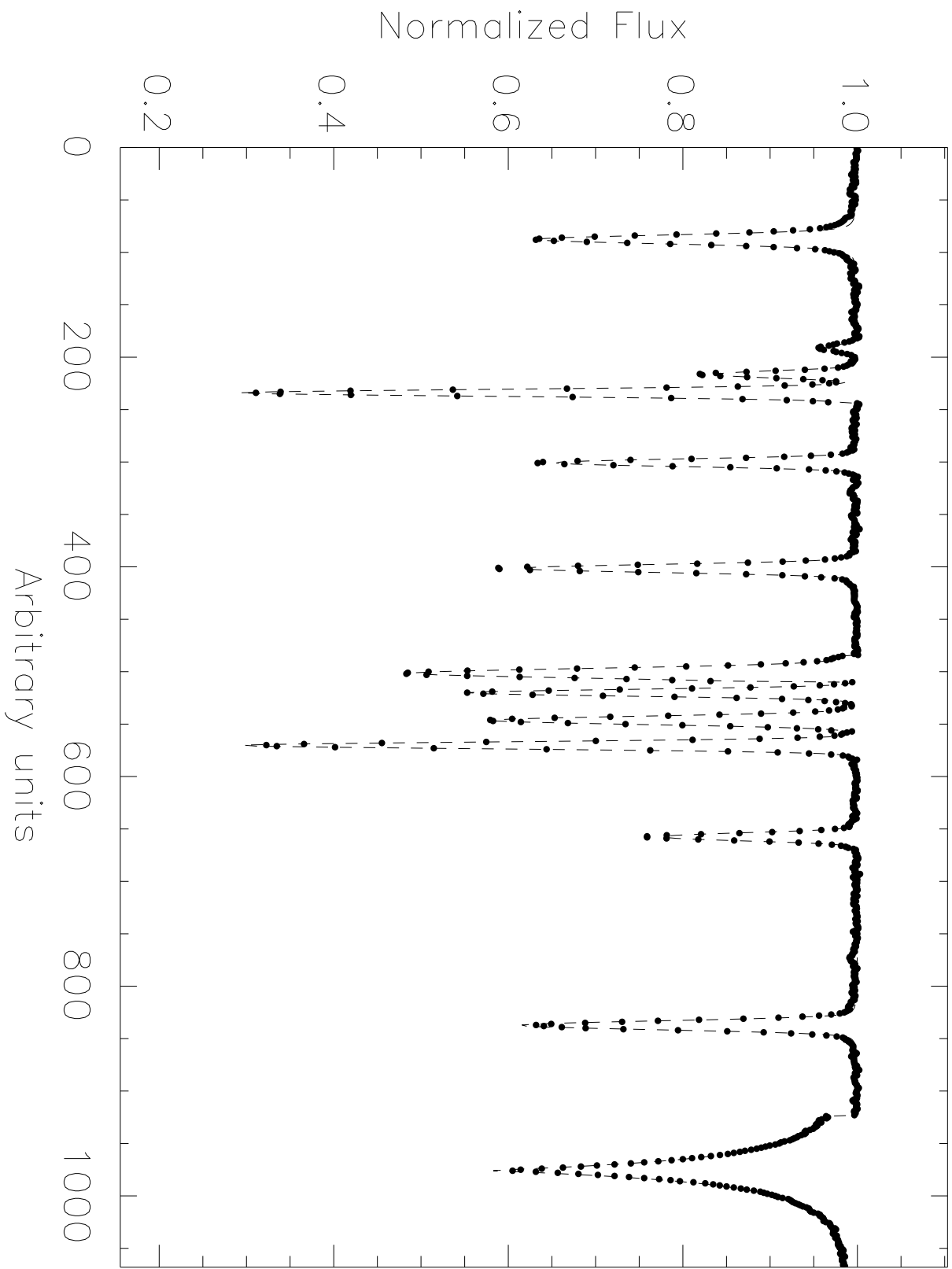
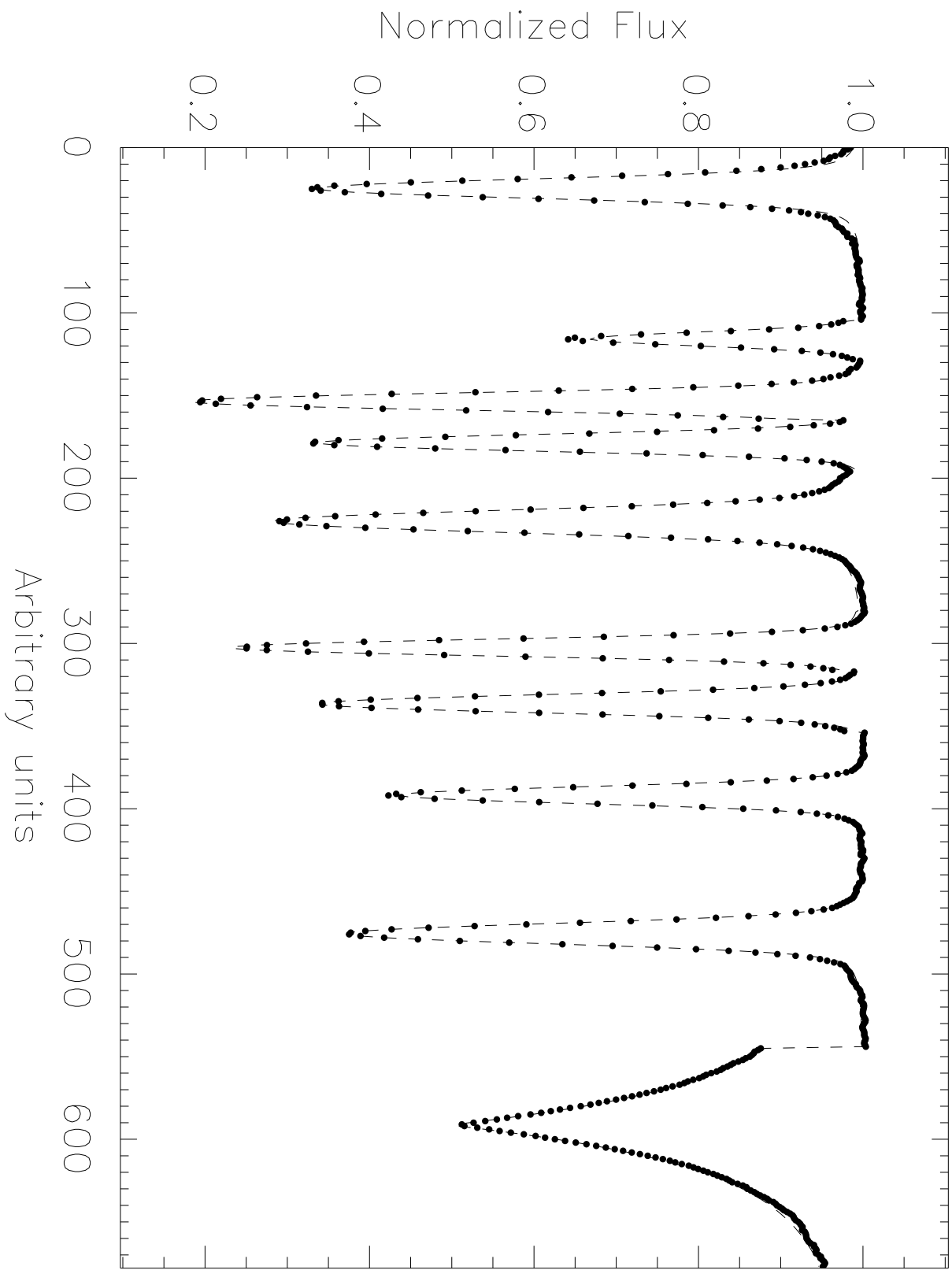
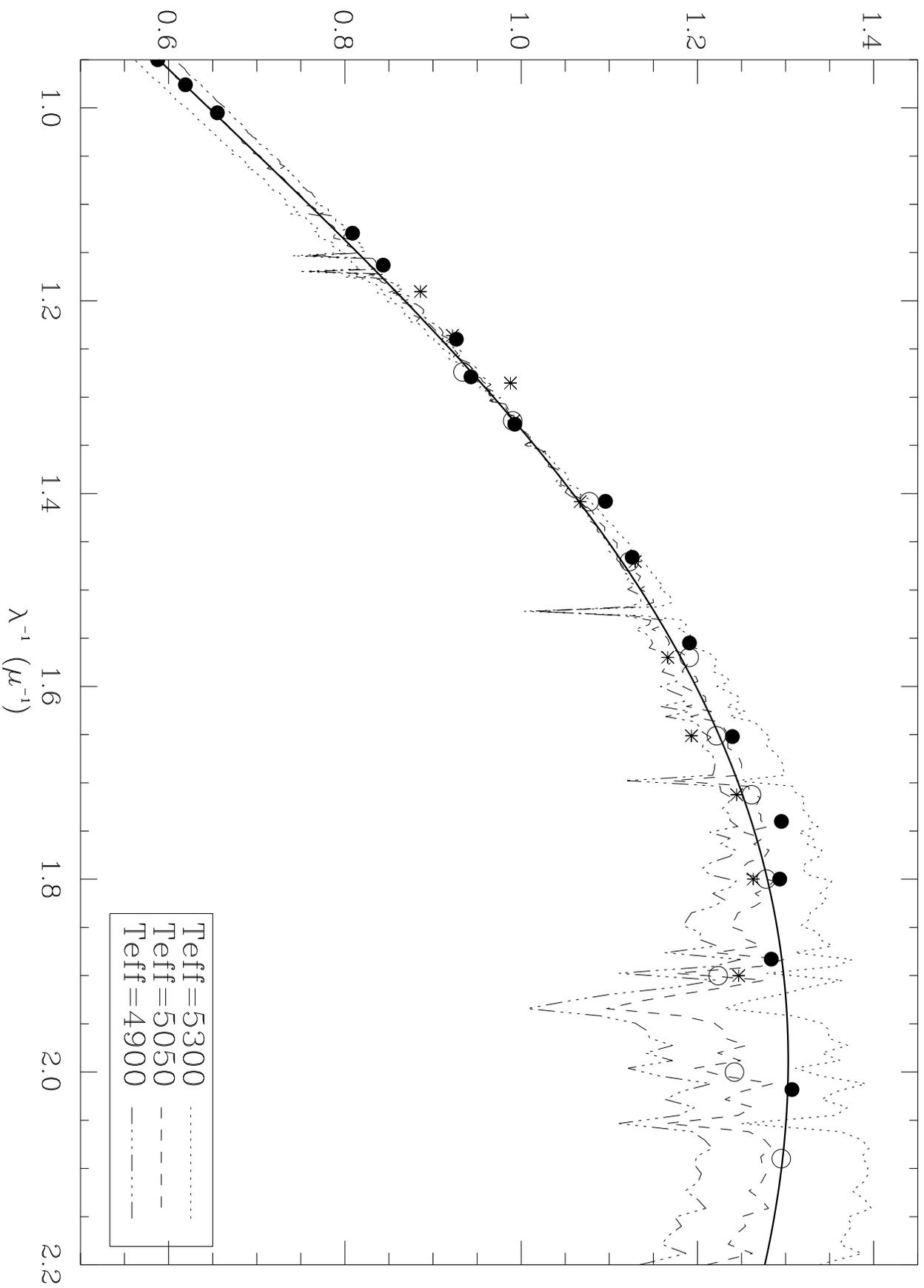
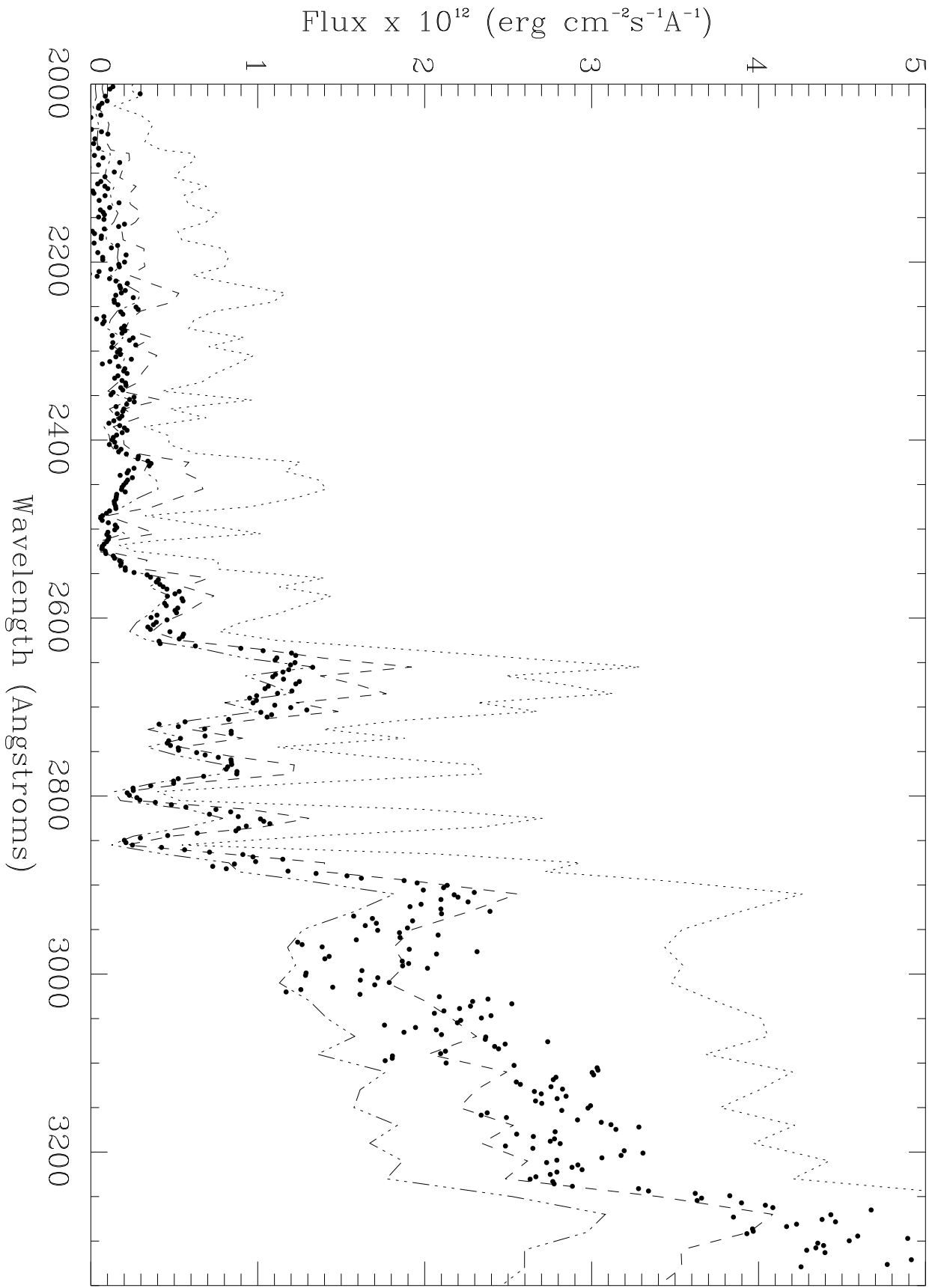
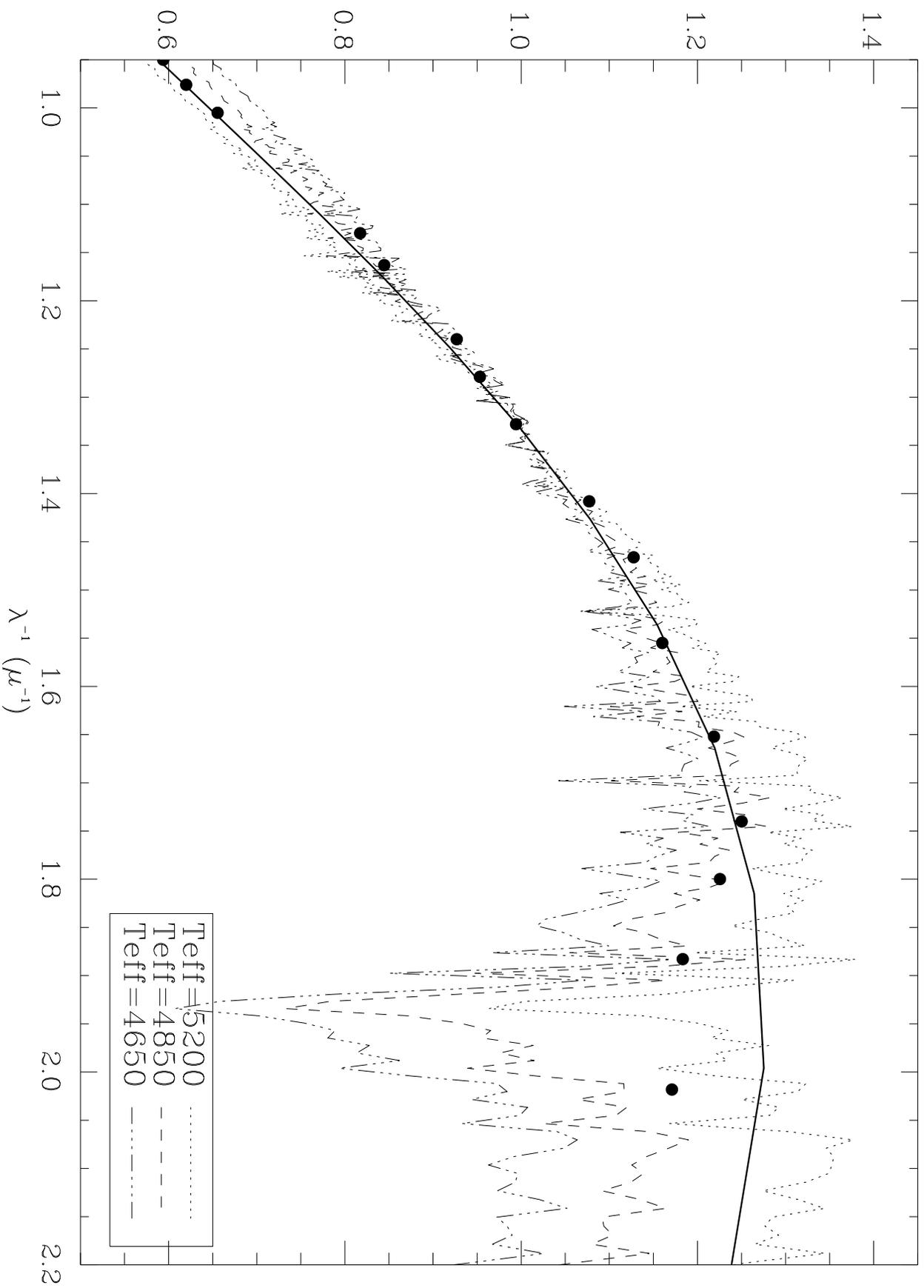
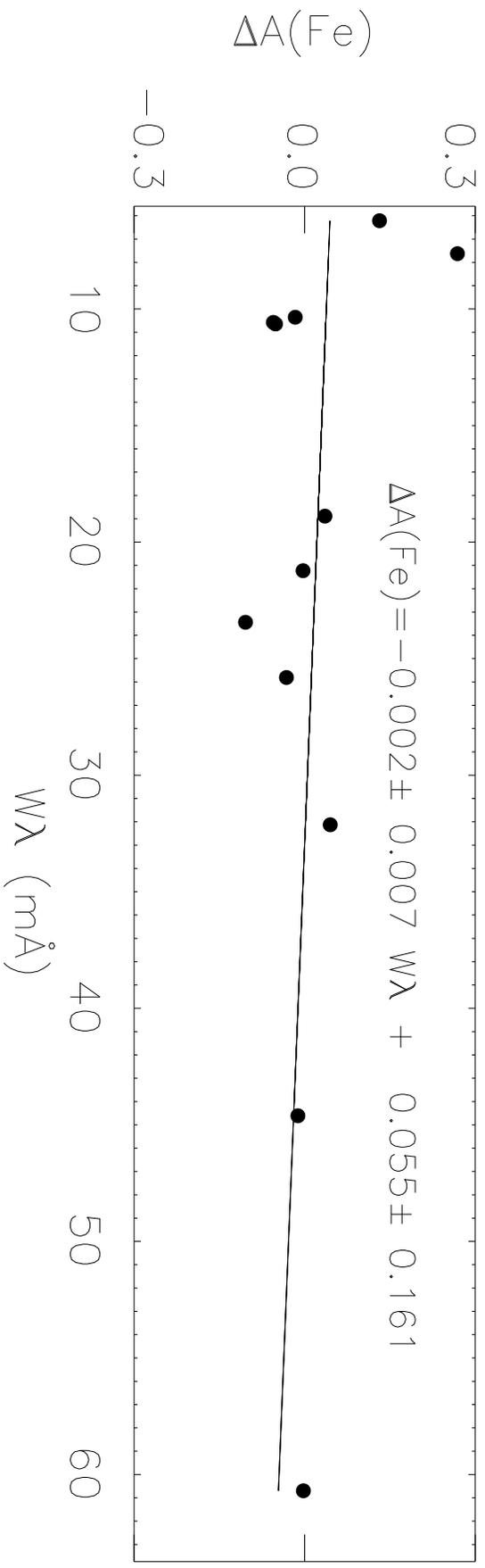
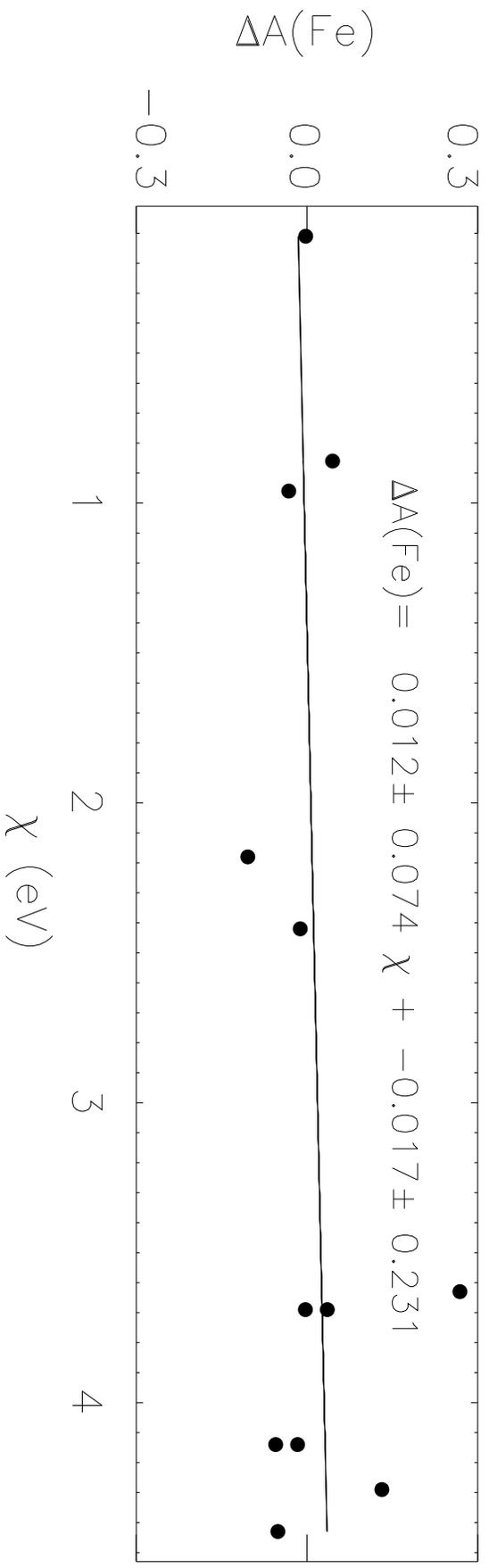
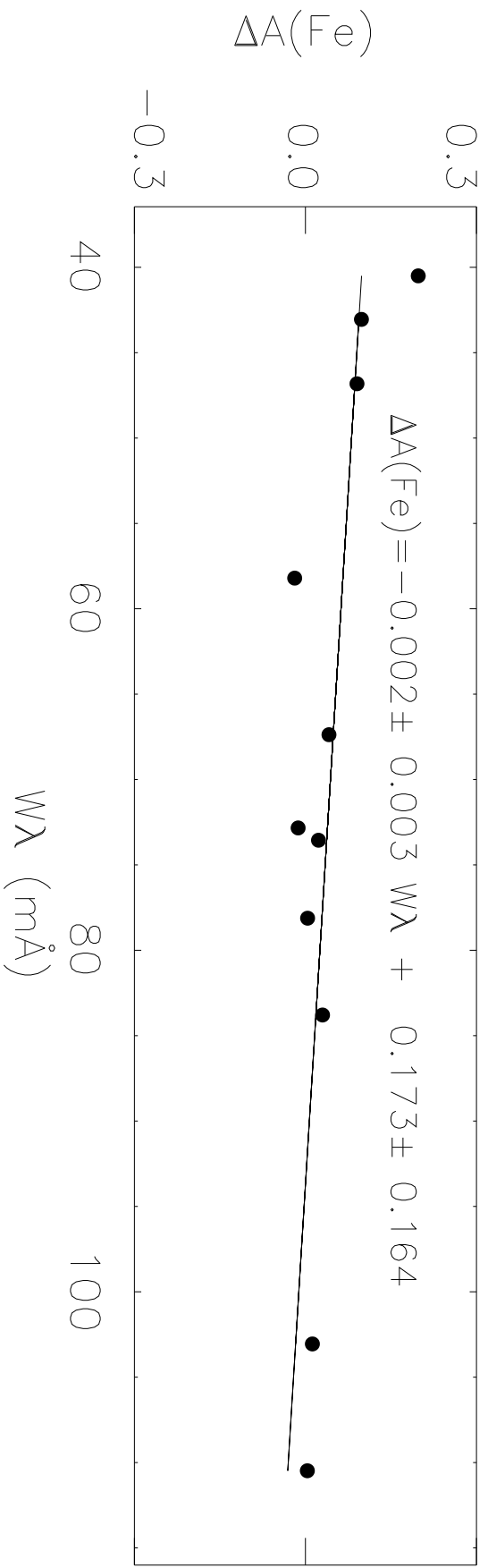
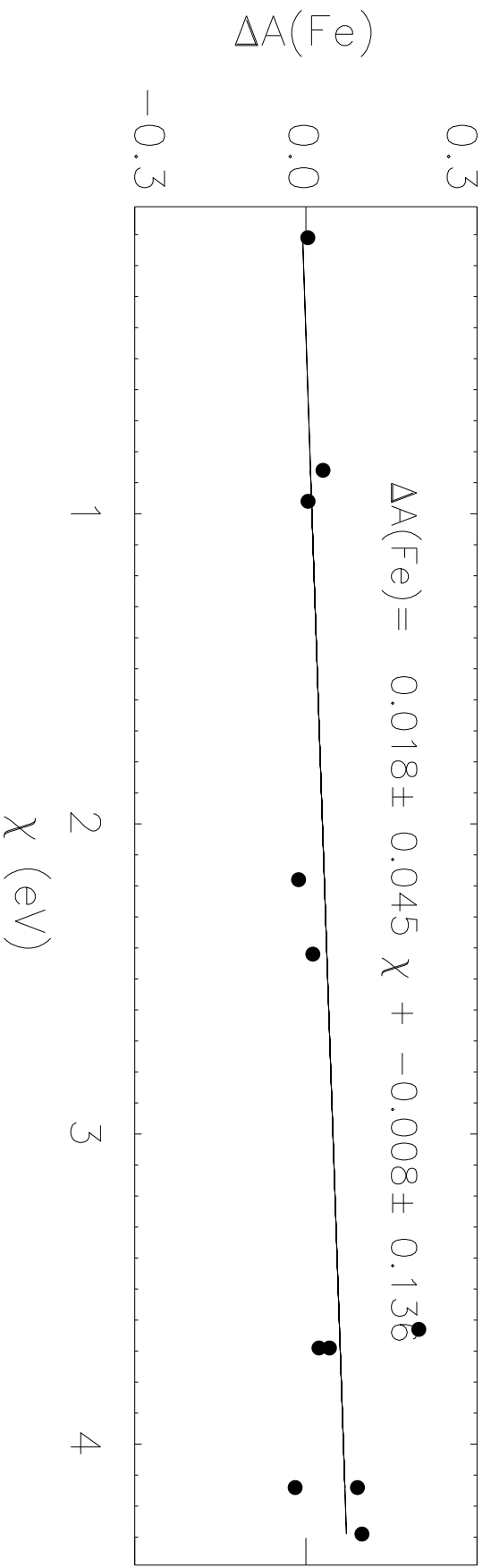
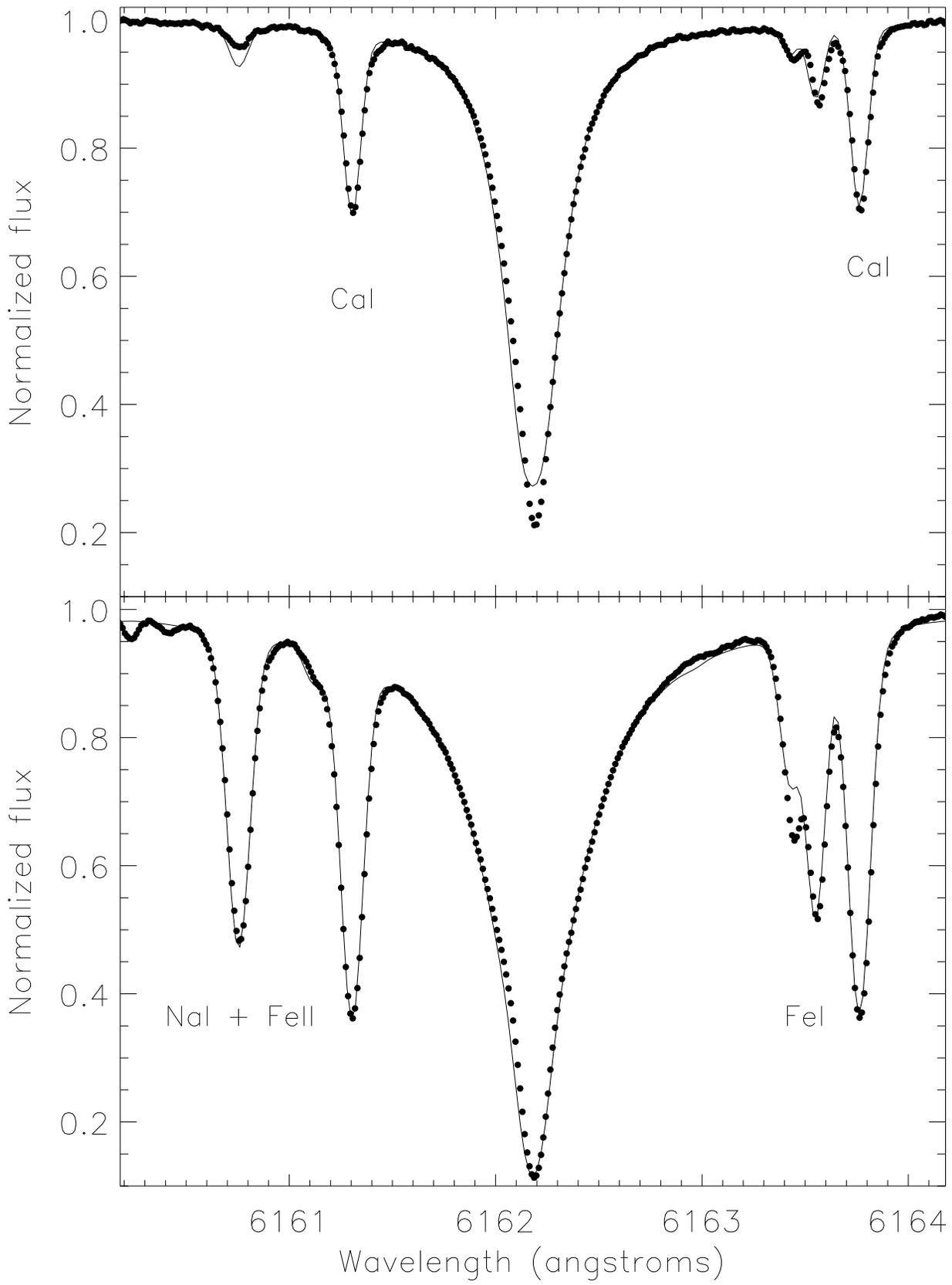
\end{document}